\documentclass[preprint,preprintnumbers,amsmath,amssymb]{revtex4}
\usepackage{amssymb,latexsym,amsmath}

\usepackage[activeacute,english]{babel}
\usepackage{fancyhdr}
\usepackage{textcomp}
\usepackage{amsfonts}
\usepackage{graphicx}
\usepackage{graphics}

\usepackage[active]{srcltx}

\newcommand{\pa}{\partial}
\newcommand{\ep}{\epsilon}

\newcommand{\la}{\lambda}

\newcommand{\om}{\omega}
\newcommand{\Om}{\Omega}

\newcommand{\rar}{\rightarrow}

\newcommand{\non}{\nonumber}

\begin{document}

\title{Two charges on a plane in a magnetic field: hidden algebra, (particular) integrability, polynomial eigenfunctions}


%
\author{A.V.~Turbiner}
\email{turbiner@nucleares.unam.mx}
\author{M.A.~Escobar-Ruiz}
\email{mauricio.escobar@nucleares.unam.mx}
\affiliation{Instituto de Ciencias Nucleares, Universidad Nacional
Aut\'onoma de M\'exico, Apartado Postal 70-543, 04510 M\'exico,
D.F., Mexico}

\date{March 10, 2013}

\begin{abstract}
The quantum mechanics of two Coulomb charges on a plane $(e_1, m_1)$ and $(e_2, m_2)$
subject to a constant magnetic field $B$ perpendicular to the plane is considered. Four integrals
of motion are explicitly indicated. It is shown that for two physically-important particular cases, namely that of two particles of equal Larmor frequencies, ${e_c} \propto \frac{e_1}{m_1}-\frac{e_2}{m_2}=0$ (e.g. two electrons) and one of a neutral system (e.g. the electron - positron pair, Hydrogen atom) at rest (the center-of-mass momentum is zero) some outstanding properties occur.
They are the most visible in double polar coordinates in CMS $(R, \phi)$ and relative $(\rho, \varphi)$ coordinate systems: (i) eigenfunctions are factorizable, all factors except one with the explicit $\rho$-dependence are found analytically, they have definite relative angular momentum, (ii) dynamics in $\rho$-direction is the same for both systems, it corresponds to a funnel-type potential and it has hidden $sl(2)$ algebra; at some discrete values of dimensionless magnetic fields $b \leq 1$, (iii) particular integral(s) occur, (iv) the hidden $sl(2)$ algebra emerges in finite-dimensional representation, thus, the system becomes {\it quasi-exactly-solvable} and (v) a finite number of polynomial eigenfunctions in $\rho$ appear. Nine families of eigenfunctions are presented explicitly.
\end{abstract}


\maketitle

A behavior of a charged particle in a constant magnetic field (the Landau problem) is well studied.
It is integrable and exactly-solvable problem, both classical and quantum, it plays a fundamental role in physics (see e.g. \cite{LL}). The problem of two charges in a magnetic field is much less studied, while the most of efforts were dedicated to explore the case when one charge is infinitely massive (one-center problem). The goal of the present paper is to consider the integrability properties (global and particular integrals, for notations, see \cite{Turbiner:2013}), the existence of algebraic structures, the presence of the hidden algebra in {\it quantum} problem of two arbitrary Coulomb charges on a plane subject to a constant uniform magnetic field.
This work is a natural continuation of \cite{ET:2013} where the {\it classical} problem of two arbitrary Coulomb charges on a plane subject to a constant uniform magnetic field was considered.
In that paper there were classified the pairs of Coulomb charges for which the special, superintegrable, closed trajectories as well as particular integrals exist. For these trajectories
the distance $\rho$ between particles remains unchanged during the evolution. Hence, $I=\rho p_{\rho}$ is the constant of motion, where $p_{\rho}$ is a component of momentum along the relative distance ${\boldsymbol \rho}$ direction.

\hskip 1cm
Let us consider a system of two non-relativistic spinless charged particles
$(e_1,\, m_1) ,\ (e_2,\, m_2)$ on a plane in a constant and uniform magnetic field
$\mathbf B=B\,\hat {\mathbf {z}}$ perpendicular to the plane. The Hamiltonian is
\begin{equation}
{\cal {\hat H}}\ =\    \frac{{({\mathbf {\hat p}_1}-e_1\,{\mathbf A_1})}^2}{2\,m_1} + \frac{{({\mathbf {\hat p}_2}-e_2\,{\mathbf A_2})}^2}{2\,m_2}  + \frac{e_1\,e_2}{\mid {\boldsymbol \rho}_1 - {\boldsymbol \rho}_2 \mid}\ ,
\label{Hcar}
\end{equation}
assuming $\hslash=c=1$, where ${\boldsymbol \rho}_{1,2}$ and ${\mathbf {\hat p}}_{{}_{1,2}}=-i\,\nabla_{{}_{1,2}}$ are the coordinate and momentum of the first (second) particle and the symmetric gauge $\mathbf A_{1,2}=\frac{1}{2}\,\mathbf B\times {\boldsymbol \rho}_{1,2}$ is chosen for vector potentials. It can be checked by direct calculation that the total Pseudomomentum
\begin{equation}
{\mathbf {\hat K}}\ =\ \mathbf {\hat p}_1+e_1\,\mathbf A_{1}+\mathbf {\hat p}_2+e_2\,\mathbf A_{2}\ ,
\label{pseudomomentumIND}
\end{equation}
is an integral, $[\, \mathbf {\hat K}, \, {\cal {\hat H}} \,]=0$ as well as the total angular momentum
\begin{equation}
\boldsymbol  {\hat L}^{\rm T } \ =\  {\boldsymbol \rho}_1 \times {\mathbf {\hat p}}_1+ {\boldsymbol \rho}_2\times {\mathbf {\hat p}}_2\,,
\label{LzT}
\end{equation}
$[\, \boldsymbol  {\hat L}^{\rm T}, \, {\cal {\hat H}} \,]=0$.

\vspace{0.2cm}

We introduce c.m.s variables
\begin{equation}
\mathbf R = \mu_1\, {\boldsymbol \rho}_1 + \mu_2\,{\boldsymbol \rho}_2 \ ,
\quad  {\boldsymbol \rho}= {\boldsymbol \rho}_1 - {\boldsymbol \rho}_2\ ,
\label{CMvar}
\end{equation}
then
\[
\mathbf {\hat P}\ =\ {\mathbf {\hat p}}_1 + {\mathbf {\hat p}}_2 \ ,
\qquad \quad \, \, {\mathbf {\hat p}}\ =\ \mu_2\,{\mathbf {\hat p}}_1 -
\mu_1\,{\mathbf {\hat p}}_2\ ,
\]
where $\mu_i=\frac{m_i}{M}$ and $M = m_1 + m_2$ is the total mass of the system, ${\mathbf {\hat P}} = -i\,\nabla_{\mathbf R},\ {\mathbf {\hat p}} = -i\,\nabla_{\boldsymbol \rho}$ are CM and relative momentum, respectively. In these coordinates the total Pseudomomentum
\begin{equation}
\mathbf {\hat K}\  =\   \mathbf {\hat P} + q\,\mathbf A_{\mathbf R} +
e_c\,\mathbf A_{{\boldsymbol \rho}}  \ ,
\label{pseudomomentum}
\end{equation}
and the total angular momentum
\begin{equation}
 \boldsymbol {\hat L}^{\rm T} \ =\ ({\mathbf R} \times {\mathbf {\hat P}}) + ({\boldsymbol \rho}\times \mathbf {\hat p})
\equiv \mathbf {\hat L} + \boldsymbol {\hat \ell} \ ,
\label{LzT2}
\end{equation}
(cf. (\ref{pseudomomentumIND}) and (\ref{LzT})),
where $\mathbf {\hat L}, \boldsymbol {\hat \ell}$ are CM angular momentum and relative angular momentum, respectively,
$$q = e_1 + e_2\ ,$$ is the total charge of the system and
\[
e_c = (e_1\,\mu_2-e_2\,\mu_1)\ =\ m_r \bigg(\frac{e_1}{m_1}-\frac{e_2}{m_2}\bigg)\ ,
\]
is a \emph{coupling} charge, $m_r=\frac{m_1\,m_2}{M}$ is the reduced mass of the system.

The operators $\mathbf {\hat K}, \boldsymbol  {\hat L}^{ \rm T }$ obey the commutation relations
\begin{equation}
\begin{aligned}
&[ {\hat K}_x,\,{\hat K}_y ] = -q\,B\,,
\\ & [ {\hat L}^{ \rm T } ,\,{\hat K}_x ] = {\hat K}_y\,,
\\ & [ {\hat L}^{ \rm T },\,{\hat K}_y ] = -{\hat K}_x\ ,
\label{AlgebraInt}
\end{aligned}
\end{equation}
hence, they span a noncommutative algebra. The Casimir operator ${\cal {\hat C}}$ of this algebra is nothing but
\begin{equation}
{\cal {\hat C}}\ =\  {\hat K}_x^2+{\hat K}_y^2-2\,q\,B\,{\hat L}^{ \rm T } \ ,
\label{Casimir}
\end{equation}

It is convenient to introduce a unitary transformation
\begin{equation}
 U = e^{-i\,e_c\,\mathbf A_{\boldsymbol \rho}\cdot \mathbf R} \ .
\label{U}
\end{equation}
Then canonical momenta transformed as,
\[
   {\mathbf {\hat Q}}=U^{-1}\,{\mathbf {\hat P}}\,U =\ {\mathbf {\hat P}}\ +\ e_c\,\mathbf A_{\boldsymbol \rho}\quad ,\quad
   {\mathbf {\hat q}}= U^{-1}\,{\mathbf {\hat p}}\, U = {\mathbf {\hat p}}\ -\
   e_c\,\mathbf A_{\mathbf R}\ .
\]
The unitary transformed Pseudomomentum (\ref{pseudomomentum}) becomes
\begin{equation}
{\mathbf {\hat K}}^{\prime}\ =\ {U}^{-1}\ {\mathbf {\hat K}}\  U\ =\ {\mathbf {\hat P}} + q\,\mathbf A_{\mathbf R}\ ,
\label{Kprime}
\end{equation}
It looks like as the Pseudomomentum of the whole, composite system of the charge $q$.
The unitary transformed Hamiltonian (\ref{Hcar}) takes the form
\begin{equation}
{\cal {\hat H}}^{\prime} \ ={U}^{-1}\ {\cal {\hat H}}\ U
\ = \ \frac{ {( \mathbf {\hat P}-q\,\mathbf A_{\mathbf R}-2\,e_c\,\mathbf A_{\boldsymbol \rho} )}^2}{2\,M}
+\frac{{({\mathbf {\hat p}}-q_\text{w}\,{\mathbf A_{\boldsymbol \rho}})}^2}{2\,m_{r}} +\frac{e_1\,e_2}{\rho}\ ,
\label{H}
\end{equation}
where $q_{\rm{w}} \equiv e_1\,\mu_2^2+e_2\,\mu_1^2$  is an effective charge (weighted total charge). It is evident, $[\, \mathbf {\hat K}^{\prime}, \, {\cal {\hat H}}^{\prime} \,]=0$\ . The eigenfunctions of ${\cal {\hat H}}$ and ${\cal {\hat H}}^{\prime}$ are related through the factor (\ref{U}),
\[
   \Psi^{\prime}\ =\ \Psi e^{ i\,e_c\,\mathbf A_{\boldsymbol \rho}\cdot \mathbf R}\ .
\]

Studying (\ref{H}) we found nothing interesting except for two special situations we will focus on, namely:

\bigskip

(i)  $e_c=0$, where separation of c.m.s. variables occurs in the Hamiltonian (\ref{H}),

\vspace{0.2cm}

(ii) $q=0$, for which components of the Pseudomomentum $\mathbf {\hat K}$ become commutative (see (\ref{AlgebraInt})).

\begin{center}
\section{Case $e_c=0$}
\end{center}

\hskip 1cm
This case appears for charges of the same sign and equal cyclotron frequency, $\frac{e_1}{m_1}=\frac{e_2}{m_2}$. The Hamiltonian $(\ref{H})$ becomes
\begin{equation}
\begin{aligned}
{\cal {\hat H}^{\prime}}& \  =\ {\cal {\hat H}} =
{\cal {\hat H}}_{R}(\mathbf {\hat P},\mathbf R) +
{\cal {\hat H}}_{\rho}(\mathbf { \hat p},\boldsymbol \rho)
\\& \equiv \frac{ {( \mathbf {\hat P}-q\,{\mathbf A}_{\mathbf R} )}^2}{2\,M} +
\frac{ {( \mathbf {\hat p}-\frac{e\,m_2}{M} {\mathbf A}_{\boldsymbol \rho} )}^2}{2\,m_r} + \frac{m_2}{m_1}\,\frac{e^2}{\rho}\ ,
\end{aligned}
\label{Hec}
\end{equation}
where CM variables are separated (see \cite{Kohn:1961} for the case of identical particles)
and $e \equiv e_1>0$\,. Here ${\cal {\hat H}}_{R}(\mathbf {\hat P},\mathbf R)$ and ${\cal {\hat H}}_{\rho}(\mathbf { \hat p},\boldsymbol \rho)$ describe CM and relative motion of two-body composite system, respectively, like it appears for field-free case.
It can be easily shown that four operators
\begin{equation}
{\cal {\hat H}}_{R}\,,\,{\cal {\hat H}}_{\rho}\,,\,{\hat {L}}_z\,,\, {\hat {l}}_z\ ,
\label{CSO}
\end{equation}
(see (\ref{LzT2}))
are mutually commuting operators spanning a commutative algebra. Hence, at $e_c=0$ the system is completely integrable. Any state is characterized by four quantum numbers.
Due to the relation
\begin{equation}
\begin{aligned}
&\mathbf{ \hat K }^2 = 2\,q\,B \,{\hat {L}}_z + 2\,M\,{\cal {\hat H}}_{R}\ ,
\end{aligned}
\label{Kx}
\end{equation}
the only one component of the Pseudomomentum is algebraically independent. Therefore, in addition to (\ref{CSO}) there exists one more algebraically independent integral, say $K_x$. Thus, in reality,  the two-body Coulomb system at $e_c=0$ in a magnetic field is globally {\it superintegrable}. These five integrals (\ref{CSO}), (\ref{Kx}) are {\it global}, they remain integrals for any state of the system.

Due to decoupling of CM and relative motion in (\ref{Hec}) the eigenfunctions are factorized
\begin{equation}
 {\cal {\hat H}} \,\Psi \ = \ (E_R+E_\rho) \, \Psi \,,\qquad \Psi=\chi(\mathbf R) \, \psi(\boldsymbol \rho)\ .
\label{Psi1}
\end{equation}
The function $\chi(\mathbf R)$ satisfies the Schr\"odinger equation
\begin{equation}
\bigg[-\frac{\nabla_R^2}{2\,M}-\frac{1}{2}\omega_c\,\hat {L}_z+\frac{M \,\om_c^2\,{R}^2}{8}\bigg]\chi(\mathbf R)\ =\ E_R \chi(\mathbf R)\ ,
\label{EqR}
\end{equation}
where $\om_c=\frac{e\,B}{m_1}= \frac{qB}{M}$ and ${\hat {L}_z}={(\mathbf R \times \mathbf {\hat P})}_z$ is CM angular momentum. This equation describes the particle of the charge $q$ and mass $M$ in a magnetic field $B$. In polar coordinates $\mathbf R=(R,\,\phi)$ the eigenfunctions and eigenvalues have a form
\begin{equation}
\begin{aligned}
& \chi \ =\  R^{| S |}\,{\rm e}^{i\,S\,\phi}\,{\rm e}^{-\frac{M\,\om_c\,R^2}{4}}\,L_N^{(| S |)}(2\,M^{-1}\,\om^{-1}_c\,R^2)\,,
\\ & E_R=\frac{\om_c}{2}(2\,N+1+|S|-S)\ ,
\end{aligned}
\label{Psiec}
\end{equation}
where $L_N^{(|S|)}$ is the associated Laguerre polynomial with index $|S|$, $N=0,\,1,\,2...$ is the principal quantum number and $S=0, \pm 1,\,\pm 2,...$ is the magnetic quantum number. There exists infinite degeneracy of $E_R$: all states at fixed $N$ and different $S \geq 0$ are degenerate. Eventually, the spectra of the Pseudomomentum reads,
\[
    K^2\ =\ q \,B\, (2\,N + 1 + |S| + S)\ .
\]
All states at fixed $N$ and different $S \leq 0$ are degenerate with respect to $K^2$.

From (\ref{Hec}) and (\ref{Psi1}) the Schroedinger equation for the relative motion can be derived
\begin{equation}
\bigg[-\frac{\nabla^2_\rho}{2\,m_r}-\frac{1}{2}\om_c\, \hat {l}_z +
\frac{m_r\,{\om_c}^2\,{\rho}^2}{8}+\frac{m_2}{m_1}\,\frac{e^2}{\rho}-E_\rho\bigg]
\psi(\boldsymbol \rho)\ =\ 0\ ,
\label{Erhoec}
\end{equation}
where $\hat {\ell}_z   =  {(\boldsymbol \rho \times \mathbf {\hat p})}_z$ is the relative angular momentum.
Equation (\ref{Erhoec}) does not admit separation of variables. However, because of vanishing the commutator $[{\cal {\hat H}}_{\rho},\,\hat {\ell}_z]=0$ an eigenfunction $\psi$ in polar coordinates $\boldsymbol \rho = (\rho, \varphi)$ has a factorized form
\begin{equation}
\begin{aligned}
&\psi(\boldsymbol \rho) \ =\  \zeta(\rho)\,\Phi(\varphi)\,,
\\ &\Phi(\varphi)  \ = \ {\rm e}^{i \,s\, \varphi}\ ,
\label{psi}
\end{aligned}
\end{equation}
where $s=0,\,\pm 1,\,\pm 2,...$ is the magnetic quantum number corresponding to the relative motion
and $\Phi(\varphi)$ is the eigenfunction of $\hat {\ell}_z$. It can be shown that the solution for $\zeta(\rho)$ has the form
\begin{equation}
\zeta_s(\rho)\ =\ \text{e}^{-\frac{m_r\,\om_c\,\rho^2}{4}}\rho^{| s|}\,p_s(\rho)\ ,
\label{zeta}
\end{equation}
where $p_s$ obeys the following equation
\begin{equation}
\bigg[-\rho\, \pa^2_\rho+(\om_c m_r \rho^2-1-2\,| s| )\,\pa_\rho+(\om_c\,\{1+|s|-s\}
-2  E_\rho)m_r\,\rho \bigg]\,p\ =\ -\ep\,p\ ,
\label{pec}
\end{equation}
where we introduce
$$\ep \equiv 2\,m_r\,e_1\,e_2=\frac{2\,m_2\,m_r\,e^2}{m_1}\ ,$$ and $\pa_\rho \equiv \frac{\pa}{\pa \rho}$. It is worth noting that similar equation for the case of two electrons was found in \cite{Taut:2000, Turbiner:1994}.
This equation can be considered as a spectral problem where $\ep$ plays a role of the spectral parameter while the energy $E_\rho$ is fixed. The parameter $\ep$ defines a strength of the Coulomb interaction. By changing variable $\rho \rar \sqrt{\om_c m_r} \rho$, Eq.({\ref{pec}}) is reduced to
\begin{equation}
  T\ p\ \equiv\ \bigg[-\rho\, \pa^2_\rho + (\rho^2  - 1 - 2|s|)\,\pa_\rho
  + \bigg(  1+|s| -s  -\frac{2\,E_\rho}{\om_c}\bigg) \rho \bigg] p
  \ =\ -\frac{\ep}{\sqrt{\om_c m_r}} p\ .
\label{P-ec}
\end{equation}
This is the basic equation we are going to study.

The operator $T$ in l.h.s. of (\ref{P-ec}) is antisymmetric: $\rho \rightarrow -\rho$ and  $T \rightarrow -T$. It implies the invariance of the equation (\ref{P-ec}): $\rho \rightarrow -\rho$ and $\ep \rightarrow -\ep$. Hence, if there exists a solution $p(\rho)$ of (\ref{P-ec}) at $\ep >0$ there must exist a solution $p(-\rho)$ with $\tilde\ep = -\ep$. It is worth mentioning the boundary conditions for (\ref{P-ec}): to assure the normalizability of $\zeta$ (\ref{zeta}), $p$ should not be too singular at the origin and should not grow faster than Gaussian at large $\rho$.

\hskip 1cm
The operator $T$ in (\ref{P-ec}) has hidden $sl_2$-Lie-algebraic structure \cite{Turbiner:1988} and the equation can be written as
\begin{equation}
   \bigg[ -{\hat J}^0_n {\hat J}^-  + {\hat J}^+_n  -
   (1 + 2 |s| + \frac{n}{2}){\hat J}^-  \bigg]p\ =\ -\frac{\ep}{\sqrt{\om_c m_r}} \,p\ ,
\label{Tec}
\end{equation}
in terms of the $sl_2$ algebra generators ${\hat J}$'s with $$n \equiv \frac{2\,E_\rho}{\om_c} + s - |s| - 1\ ,$$ which defines the spin of representation, see (\ref{generators}). For fixed $n$ the energy of the relative motion is
\begin{equation}
 E_\rho = \frac{\om_c}{2}\,(n+1+|s|-s)\ .
\label{Erho}
\end{equation}
Thus, there exists infinite degeneracy of $E_\rho$: all states at fixed $n$ and different relative magnetic quantum numbers $s \geq 0$ are degenerate.
Eventually, solutions for $p$ depends on $|s|$, $p = p_{|s|} (\sqrt{\om_c m_r} \rho)$.

A hidden algebraic structure occurs (the underlying idea behind quasi-exactly solvability) at nonnegative integer $n$: the hidden $sl_2$ algebra appears in finite-dimensional representation and the problem (\ref{Tec}) possesses $(n+1)$ eigenfunctions $p_{n,i}, \ i=1,\ldots (n+1)$ in the form of a polynomial of the $n$th power. It implies the explicit (algebraic) quantization of $\ep$, since $\ep$'s are roots of algebraic secular equation, and thus, a quantization of the dimensionless parameter
\begin{equation}
 \la \equiv \frac{\ep^2}{\om_c\,m_r}\ =\  \frac{B_0}{B} \equiv \frac{1}{b}\ ,\quad B_0=4\,m_r M\, \frac{e_1^2 e_2^2}{e_1+e_2}\ ,
\label{lambdaec}
\end{equation}
which we introduce for convenience, here $B_0$ has a meaning of the characteristic magnetic field. Hence, for a given system with fixed masses and charges at $e_c=0$, there exists an infinite, discrete set of values of the magnetic field $B$ for which the exact, analytic solutions of the Schr\"odinger equation occur. The parameter $\ep$ takes $[\frac{n+1}{2}]$ positive values and $[\frac{n+1}{2}]$ symmetric negative values, and zero value for even $n$, for which the problem (\ref{Tec}) possesses polynomial solutions \footnote{[$a$] means integer part of $a$.}. From physical point of view, only $\ep > 0$ are admitted. It is worth noting that for even $n$, always there is a solution at $\la_n=0$, hence $\ep=0$. It corresponds to either non-normalizable wavefunction, $\om_c=0$, or vanishing Coulomb interaction, $\la=0$. We should skip such a solution as unphysical when we study the spectra (see below). The interesting fact should be emphasized: for given $n$ the $[\frac{n+1}{2}]$ physically admitted functions $p_n$ have a number of nodes at $\rho >0$ varying from zero up to $[\frac{n-1}{2}]$.

Below we present analytical results for $\la_n$ ($\ep=\sqrt{\om_c\,m_r\,\la_n}$) and the corresponding eigenfunctions $p_n$ up to $n=8$.

\vspace{0.2cm}

$\bullet$\ $n=0$

\begin{equation}
\begin{aligned}
& \la_0 \ =\  0\ .
\\ & p_0 \ = \  1\ .
\label{n=0ec}
\end{aligned}
\end{equation}
It corresponds to either vanishing Coulomb interaction or infinite magnetic field $B$, thus, $\ep=0$. In the case of two electrons, for $N=n=0$ the total wavefunction $\Psi$ (\ref{Psi1}) coincides exactly with the wavefunction proposed by R.~Laughlin for the quantum Hall effect \cite{Laughlin:1983}.

$\bullet$\  $n=1$

\begin{equation}
\begin{aligned}
& \la_1 \  =  \ 1+2\,|s|\ ,
\\ & p_1(\rho) \ = \  1 + \frac{\sqrt{\la_1}}{1 + 2\,|s|}\,\rho\ .
\label{n=1ec}
\end{aligned}
\end{equation}
This solution corresponds to the ground state: $p_1$ has no node at $\rho \geq 0$. It appears
at magnetic field
\[
     b_1\ =\ \frac{1}{1+2\,|s|} \ ,
\]
if measured in the characteristic magnetic field $B_0$, see (\ref{lambdaec}). It takes the maximal value $b_{1,max}\ =\ 1$ at $s=0$. It seems $b=1$ is the maximal magnetic field for the analytic solution exists.

$\bullet$\ $n=2$

\begin{equation}
\begin{aligned}
\\ & \la_2  \  = \  6+8\,|s|\,. \\
& p_2(\rho)\ =\ 1 + \frac{\sqrt{\la_2}}{1 + 2\,|s|}\,\rho
+ \frac{\la_2-2-4\,|s|}{4+12\,|s|+8\,|s|^2}\,\rho^2\ .
\label{n=2ec}
\end{aligned}
\end{equation}
Since $p_2$ has no nodes at $\rho \geq 0$, it corresponds to ground state, $\ep_2 > 0$. It appears
at magnetic field
\[
     b_2\ =\ \frac{1}{6+8\,|s|} \ .
\]
It takes the maximal value $b_{2,max}\ =\ \frac{1}{6}$ at $s=0$.

$\bullet$\  $n=3$

There exist two different polynomial solutions $p_{3,j}, j=1,2,$ in a form of the polynomial of the 3rd degree with $\ep_3 >0$ with no and one (positive) node, respectively. They correspond to
\begin{equation}
\la_{3,j} \ =\ 10 + 10\,|s|-{(-1)}^{j} \sqrt{73 + 128|s| + 64{|s|}^2}\ ,\quad j=1,2\ ,
\end{equation}
\begin{equation}
\label{n=3ec}
   p_{3,j}\ =\ 1 + \frac{\sqrt{ \la_{3,{j}}}}{1 + 2\,|s|}\,\rho +
   \frac{ \la_{3,{j}}-3-6\,|s|}{4+12\,|s|+8\,|s|^2}\,\rho^2
   + \frac{\sqrt{ \la_{3,j} } \,( \la_{3,j}-11-14\,|s|)}{12\,(1+|s|)(1+2\,|s|)(3+2\,|s|)}\,\rho^3    \ .
\end{equation}
These eigenfunctions appear at magnetic fields
\[
     b_{3,j}\ =\ \frac{1}{10 + 10\,|s|-{(-1)}^{j} \sqrt{73 + 128|s| + 64{|s|}^2}} \ ,
\]
respectively. For the case of the ground state, $j=1$, it takes the maximal value
$b_{3,1,max}~=~\frac{1}{10 + \sqrt{73}}$ at $s=0$.

$\bullet$\ $n=4$

There exist two different polynomial solutions $p_{4,j},\ j=1,2,$ in a form of the polynomial of the 4th degree with $\ep_4 >0$ with no and one (positive) node, respectively. They correspond to
\begin{equation}
\la_{4,j} \  = \ 25 + 20\,|s| +{(-1)}^{j+1} 3\sqrt{33 + 40|s| + 16{|s|}^2}\ ,
\end{equation}
\begin{equation}
\begin{aligned}
  p_{4,j} &=\ 1 + \frac{\sqrt{\la_{4,j}}}{1 + 2\,|s|}\,\rho + \frac{\la_{4,j} -4-8\,|s|}{4+12\,|s|+8\,|s|^2}\,\rho^2
   + \frac{\sqrt{\la_{4,j}}\,(\la_{4,j} -16-20\,|s|)}{12\,(1+|s|)(1+2\,|s|)(3+2\,|s|)}\,\rho^3  \\ &
   +\frac{\la_{4,j}^2-\la_{4,j}\,(34+32\,|s|)+24\,(3+8\,|s|+4\,|s|^2)}{96\,(1+|s|)(2+|s|)(1+2\,|s|)(3+2\,|s|)}\,\rho^4  \ .
\end{aligned}
\label{n=4ec}
\end{equation}
These eigenfunctions appear at magnetic fields
\[
     b_{4,j}\ =\ \frac{1}{25 + 20\,|s| +{(-1)}^{j+1} 3\sqrt{33 + 40|s| + 16{|s|}^2}} \ ,
\]
respectively. For the case of the ground state, $j=1$, it takes the maximal value
$b_{4,1,max}~=~\frac{1}{25 + 3 \sqrt{33}}$ at $s=0$.

$\bullet$\ $n=5$

In this case there exist three different polynomial solutions $p_{5,j},\ j=1,2,3,$ in a form of the polynomial of the 5th degree with $\ep_5 >0$ with no, one and two (positive) nodes, $j=1,3,2$~, respectively. They correspond to
\begin{equation}
 \la_{5,j} = \frac{4}{3}\,\sqrt{1251 + 448|s|(3 + |s|)}\cos\bigg(\frac{\theta+2(j-1)\,\pi}{3}\bigg)+\frac{35}{3}\,(3+2| s|)\ ,
\end{equation}
where
\[
\theta = \cos^{-1}\bigg(\frac{20\,( 3 + 2|s|)(531+384| s|+128{| s|}^2)}{{[1251 + 448| s|(3 + |s|)]}^{\frac{3}{2}}}\bigg)\ .
\]
and
\begin{equation}
\begin{aligned}
 p_{5,j} &=\ 1 + \frac{\sqrt{\la_{5,j}}}{1 + 2\,|s|}\,\rho +\ \frac{\la_{5,j}-5-10\,|s|}{4+12\,|s|+8\,|s|^2}\,\rho^2
 + \frac{\sqrt{\la_{5,j}}\,(\la_{5,j} -21-26\,|s|)}{12\,(1+|s|)(1+2\,|s|)(3+2\,|s|)}\ \rho^3  \\&
 +\frac{\la_{5,j}^2-4\,\la_{5,j}\,(12+11\,|s|)+45\,(3+8\,|s|+4\,|s|^2)}{96\,(1+|s|)(2+|s|)(1+2\,|s|)
 (3+2\,|s|)}\ \rho^4
\\ &+ \frac{\sqrt{\la_{5,j}}\,[ \la_{5,j}^2-\la_{5,j}\,(80+60\,|s|)+807+1528\,|s|+596\,|s|^2   ]}{480\,(1+|s|)(2+|s|)(1+2\,|s|)(3+2\,|s|)(5+2\,|s|)} \ \rho^5 \ .
\end{aligned}
\label{n=5ec}
\end{equation}
These eigenfunctions appear at magnetic fields
\[
     b_{5,j}\ =\ \frac{1}{\la_{5,j}} \ ,
\]
respectively.

$\bullet$\  $n=6$

There exist three different polynomial solutions $p_{6,j},\ j=1,2,3,$ in a form of the polynomial of the 6th degree with $\ep_6 >0$ with no, one and two (positive) nodes, $j=1,3,2$\ , respectively. They correspond to
\begin{equation}
  \la_{6,j}  = \frac{4}{3}\,\sqrt{3211 + 392|s| (7 +2|s|)}
  \cos\bigg(\frac{\theta+2(j-1)\,\pi}{3}\bigg)+\frac{28}{3}\,(7+4| s|)\ ,
\end{equation}
where
\begin{equation}
\theta =\cos^{-1}\bigg(\frac{8\,(7 + 4|s|)(1939 + 1001| s|+286{|s|}^2)}{{[3211 + 392|s|(7+2| s|)]}^{\frac{3}{2}}}\bigg)\ . \non
\end{equation}
\begin{equation}
\begin{aligned}
&  p_{6,j} = 1 + \frac{\sqrt{ \la_{6,j}}}{1 + 2\,|s|}\,\rho + \frac{\la_{6,j} -6-12\,|s|}{4+12\,|s|+8\,|s|^2}\ \rho^2
   + \frac{\sqrt{\la_{6,j}}\,(\la_{6,(j)} -26-32\,|s|)}{12\,(1+|s|)(1+2\,|s|)(3+2\,|s|)}\,\rho^3  \\ &
   +\frac{\la_{6,j}^2-\,\la_{6,(j)} \,(62+56\,|s|)+72\,(3+8\,|s|+4\,|s|^2)}{96\,(1+|s|)(2+|s|)(1+2\,|s|)(3+2\,|s|)}\,\rho^4
   \\ & + \frac{\sqrt{\la_{6,j}}\,[ \la_{6,j}^2-\la_{6,j} \,(110+80\,|s|)+24\,(61+114\,|s|+44\,|s|^2)  ]}{480\,(1+|s|)(2+|s|)(1+2\,|s|)(3+2\,|s|)(5+2\,|s|)}\,\rho^5
   \\ & + \frac{\la_{6,j}^3- 2\la_{6,j}^2\,(80+50|s|) +4\la_{6,j}[1141+2|s|(847+272|s|)]-720(1+2\,|s|)(3+2|s|)(5+2|s|)    }{5760\,(1+|s|)(2+|s|)(3+|s|)(1+2\,|s|)(3+2\,|s|)(5+2\,|s|)}\ \rho^6  \ .
\end{aligned}
\label{n=6ec}
\end{equation}
These eigenfunctions appear at magnetic fields
\[
     b_{6,j}\ =\ \frac{1}{\la_{6,j}} \ ,
\]
respectively.

$\bullet$\ $n=7$

There exist four different polynomial solutions $p_{7,j},\ j=1,2,3,4$ in a form of the polynomial of the 7th degree with $\ep_7 >0$ with no, one, two and three (positive) nodes, respectively. They correspond to
\begin{equation}
\begin{aligned}
&      \la_{7,1} = 42\,(2 + |s|)\, + \sqrt{z_1} + \sqrt{z_2} + \sqrt{z_3}\,,
\\ &  \la_{7,2} = 42\,(2 + |s|)\, + \sqrt{z_1} - \sqrt{z_2} - \sqrt{z_3}\,,
\\ &  \la_{7,3} = 42\,(2 + |s|)\, - \sqrt{z_1} - \sqrt{z_2} + \sqrt{z_3}\ ,
\\ &  \la_{7,4} = 42\,(2 + |s|)\, - \sqrt{z_1} + \sqrt{z_2} - \sqrt{z_3}\ ,
\end{aligned}
\end{equation}
where
\begin{equation}
\begin{aligned}
&  z_i=\sqrt{ 12\,[A_1 + 896 | s|(4 + |s|)\{281+32| s|(4 + |s|)\}]}\cos\bigg(\frac{\theta + 2\,i\,\pi}{3}\bigg) + f_s\ ,\ i=1,2,3
\\ &  f_s=2287+448|s|(4+|s|)\,,
\\ &  \theta = \cos^{-1} \bigg[  \frac{9\sqrt{3}[A_2 + 64|s|(4 + |s|)(A_3 + 64|s|(4 + |s|)\{843 + 64|s|(4 + |s|)\})   ]}{{[A_1+896 | s|(4+|s|)\{281 + 32|s|(4 + |s|)\} ]}^{\frac{3}{2}}}  \bigg]\,,
\\ &  A_1 = 571527\,,\quad A_2 = 24416241\,,\quad A_3 = 246083\ . \non
\end{aligned}
\end{equation}
\begin{equation}
\begin{aligned}
&  p_{7,j} = 1 + \frac{  \sqrt{ \la_{7,j}}   }{1 + 2\,|s|}\ \rho
+ \frac{\la_{7,j} -7-14\,|s|}{4+12\,|s|+8\,|s|^2}\ \rho^2
   + \frac{\sqrt{\la_{7,j} }\,(\la_{7,j} -31-38\,|s|)}{12\,(1+|s|)(1+2\,|s|)(3+2\,|s|)}\ \rho^3  \\ &
   +\frac{\la_{7,j}^2-2\,\la_{7,j} \,(38+34\,|s|)+315+840\,|s|+420\,|s|^2}{96\,(1+|s|)(2+|s|)(1+2\,|s|)(3+2\,|s|)}\ \rho^4
   \\ & + \frac{\sqrt{\la_{7,j} }\,[ \la_{7,j}^2-\la_{7,j} \,(140+100\,|s|)+(2299+4264\,|s|+1636\,|s|^2)  ]}{480\,(1+|s|)(2+|s|)(1+2\,|s|)(3+2\,|s|)(5+2\,|s|)}\ \rho^5\ +
   \\ &  \frac{\la_{7,j}^3- 5\la_{7,j}^2\,(43+26|s|) +\la_{7,j}[7999+4|s|(2911+919|s|)]-1575(1+2\,|s|)(3+2|s|)(5+2|s|)    }{5760\,(1+|s|)(2+|s|)(3+|s|)(1+2\,|s|)(3+2\,|s|)(5+2\,|s|)}\ \rho^6
   \\ & + \frac{\sqrt{\la_{7,j}}\,(\la_{7,j}^3-7\,\la_{7,j}^2\,(41+22\,|s|) +28\,\la_{7,j}\,|s|\,(793+217\,|s|)+\la_{7,j}18079 + F_s) }{40320\,(1+|s|)(2+|s|)(3+|s|)(1+2\,|s|)(3+2\,|s|)(5+2\,|s|)(7+2\,|s|)}\ \rho^7                      \ ,
\end{aligned}
\label{n=7ec}
\end{equation}
where
$$F_s=-189153-6 \,|s|\,(72439+46138\,|s|+8644\,|s|^2) \ .$$
These eigenfunctions appear at magnetic fields
\[
     b_{7,j}\ =\ \frac{1}{\la_{7,j}} \ ,
\]
respectively.

$\bullet$\ $n=8$

There exist four different polynomial solutions $p_{8,j},\ j=1,2,3,4$ in a form of the polynomial of the 8th degree with $\ep_8 >0$ with no, one, two and three (positive) nodes, respectively. They correspond to
\begin{equation}
\begin{aligned}
   &  \la_{8,1} = 15\,(9 + 4|s|) + \sqrt{z_1} + \sqrt{z_2} + \sqrt{z_3}\,,
\\ &  \la_{8,2} = 15\,(9 + 4|s|) + \sqrt{z_1} - \sqrt{z_2} - \sqrt{z_3}\,,
\\ &  \la_{8,3} = 15\,(9 + 4|s|) - \sqrt{z_1} - \sqrt{z_2} + \sqrt{z_3}\,,
\\ &  \la_{8,4} = 15\,(9 + 4|s|) - \sqrt{z_1} + \sqrt{z_2} - \sqrt{z_3}\ .
\end{aligned}
\end{equation}
where
\begin{equation}
\begin{aligned}
& z_i = \sqrt{ 12\,[A_1 + 4528|s|(9 + 2|s|)\{ 97 + 4|s|(9 + 2| s|)\}]}\cos\bigg(\frac{\theta + 2\,i\,\pi}{3}\bigg) + f_s\ ,\ i=1,2,3
\\ &  f_s = 4671 + 344|s|(9+2|s|)\,,
\\ &  \theta = \cos^{-1} \bigg[  \frac{9\sqrt{3}[A_2 + 8|s|(9 + 2|s|)(A_3 + 1976|s|(9 + 2|s|)\{291 + 8|s|(9 + 2|s|)\})   ]}{{[A_1 + 4528|s|(9+2|s|)\{97+4| s|(9+2| s|)\} ]}^{\frac{3}{2}}}  \bigg]\,,
\\ &   A_1 = 2750247\,,\quad A_2 = 246596481\,,\quad A_3 = 7243319\ .\non
\label{n=8ecc}
\end{aligned}
\end{equation}
\begin{equation}
\begin{aligned}
&  p_{8,j} = 1 + \frac{  \sqrt{ \la_{8,j} }   }{1 + 2\,|s|}\,\rho
+ \frac{\la_{8,j} -8-16\,|s|}{4+12\,|s|+8\,|s|^2}\,\rho^2
   + \frac{\sqrt{\la_{8,j} }\,(\la_{8,j} -36-44\,|s|)}{12\,(1+|s|)(1+2\,|s|)(3+2\,|s|)}\,\rho^3  \\ &
   +\frac{\la_{8,j}^2-\la_{8,j} \,(90+80\,|s|)+144(3+8\,|s|+4\,|s|^2)}{96\,(1+|s|)(2+|s|)(1+2\,|s|)(3+2\,|s|)}\ \rho^4
\\ & + \frac{   \sqrt{\la_{8,j} }\,[ \la_{8,j}^2-\la_{8,j}
  \,(170+120\,|s|)+(3312+6112\,|s|+2336\,|s|^2)  ]}{480\,(1+|s|)(2+|s|)(1+2\,|s|)(3+2\,|s|)(5+2\,|s|)}\ \rho^5\ +
\\ &  \frac{\la_{8,j}^3- 10\la_{8,j}^2\,(27+16|s|)
  +8\,\la_{8,j}[1539+|s|(2214+692|s|)]-2880(1+2\,|s|)(3+2|s|)(5+2|s|)    }{5760\,(1+|s|)(2+|s|)(3+|s|)(1+2\,|s|)(3+2\,|s|)(5+2\,|s|)}\ \rho^6
\\ & + \frac{ \sqrt{\la_{8,j}}\,(\la_{8,j}^3-14\,\la_{8,j}^2\,(27+14\,|s|)\ +\
  7\,\la_{8,j}\,|s|\,(657+176\,|s|)+\la_{8,j}30672 + F_s) }{40320\,(1+|s|)(2+|s|)(3+|s|)(1+2\,|s|)(3+2\,|s|)(5+2\,|s|)(7+2\,|s|)}\ \rho^7
\\ &\ +\ \frac{\la_{8,j}^4-28\la_{8,j}^3(17+8|s|)+4\la_{8,j}^2[14283+224|s|(67+16|s|)] +
  16\la_{8,j}G_s + D_s}   {645120\,(1+|s|)(2+|s|)(3+|s|)(4+|s|)(1+2\,|s|)(3+2\,|s|)(5+2\,|s|)(7+2\,|s|)}\ \rho^8 \ .
\end{aligned}
\label{n=8ec}
\end{equation}
where
\begin{equation}
\begin{aligned}
& F_s=-400896-576 \,|s|\,(1583+1000\,|s|+186\,|s|^2)\ ,
\\ &G_s=-100467 - 4 |s| (46755 + 25226 |s| + 4096 |s|^2) \ ,
\\ &D_s=40320 (1 + 2 s) (3 + 2 s) (5 + 2 s) (7 + 2 s)\ . \non
\end{aligned}
\end{equation}
These eigenfunctions appear at magnetic fields
\[
     b_{8,j}\ =\ \frac{1}{\la_{8,j}} \ ,
\]
respectively.

\subsection{Particular integral}

\hskip 1cm
Let take the Euler-Cartan operator
\[
   i^0_n\ =\ \rho {\pa_\rho} - n\ ,
\]
and form the operator
\begin{equation}
\label{INT}
      i_n(\rho)\ =\ \prod_{j=0}^n (\rho {\pa_\rho} + j)\ .
\end{equation}
This operator has a property of annihilator
\[
  i_n(\rho):\ {\cal P}_{n} \ \mapsto \ 0\ ,
\]
where ${\cal P}_{n}$ is the linear space of polynomials in $\rho$ of degree not higher than $n$, (\ref{Pn}). It is evident that for the operator $T(n)$ at integer $n$ (see (\ref{P-ec}), (\ref{Tec}))
\[
  [T(n), i_n(\rho)]: {\cal P}_{n} \ \mapsto \ 0\ .
\]
Hence, $i_n(\rho)$ is the particular integral with ${\cal P}_{n}$ as the invariant subspace. The eigenfunctions $p_{n, k}\ , \ k=1,\ldots (n+1)$ of (\ref{P-ec}), (\ref{Tec})) are the zero modes of $i_n(\rho)$. Taking the gauge rotated $i_n(\rho)$ with the factor
$\zeta^{(0)} = \text{e}^{-\frac{m_r\,\om_c\,\rho^2}{4}}\rho^{|s|}\ $ (see (\ref{zeta})),
\begin{equation}
\label{INT-H}
      \zeta^{(0)}\ i_n\ (\rho) (\zeta^{(0)})^{-1}\ =\ \prod_{j=0}^n (\rho {\cal D}_{\rho} + j) \equiv {\cal I}_n\ (\rho)
      \ ,
\end{equation}
where $${\cal D}_{\rho} = \pa _{\rho} + \frac{m_r\,\om_c\,\rho}{2} - \frac{|s|}{\rho}\ ,$$ is the covariant derivative, we arrive at
\[
  [{\cal {\hat H}}_{\rho}(\mathbf { \hat p},\boldsymbol \rho)\ ,\
  {\cal I}_n(\rho)]: {\cal V}_{n} \ \mapsto \ 0\ ,
\]
at $b=b_n$, or equivalently at  $\om_c=({\om_c})_n$.
Hence, ${\cal I}_n(\rho)$ is the particular integral with ${\cal V}_{n}\ =\ \zeta^{(0)} {\cal P}_n$ as the invariant subspace.

\hskip 1cm
In classical limit the particular integral ${\cal I}_n(\rho)$ becomes the classical particular integral $I_n(i\ \rho p_{\rho})$ (see \cite{ET:2013}). The latter becomes the constant of motion on the special periodic trajectories. It manifests a connection between quasi-exact-solvability in quantum mechanics and special trajectories in classical mechanics \cite{Turbiner:2013}. Eventually, the planar Coulomb system with $e_c=0$ in a magnetic field $B$, both classical and quantum, is completely integrable and superintegrable with five global integrals at any $B$ and one more particular integral for a certain values of $B$.

\vskip 0.5cm

\section{Neutral Case, $q=0$}

\hskip 1cm
Let us consider a neutral system, $e_1=-e_2\equiv e$. The unitary-transformed Hamiltonian $(\ref{H})$ becomes
\begin{equation}
{\cal {\hat H}}^{\prime} \ =\  \frac{{(\mathbf {\hat P}-e\,
{\mathbf B \times {\boldsymbol \rho}} )}^2}{2\,M}\
+\ \frac{{({\mathbf {\hat p}} - e(\mu_2-\mu_1)\,{\mathbf A_{\boldsymbol \rho}})}^2}{2\,m_{r}}\ -\ \frac{e^2}{\rho}\ .
\label{Hq}
\end{equation}
The variables in $(\ref{Hq})$ are not separated. In order to proceed we consider a bispectral problem for the unitary transformed Hamiltonian and Pseudomomentum
\begin{equation}
 {\cal {\hat H}}^{\prime} \,\Psi_{{}_{\mathbf K}}^{\prime}\ =\ E \,\Psi_{{}_{\mathbf K}}^{\prime}\,, \qquad
 \mathbf {\hat K}^{\prime}\,\Psi_{{}_{\mathbf K}}^{\prime}\ =\ \bf K\,\Psi_{{}_{\mathbf K}}^{\prime}
\label{biespectral}\ .
\end{equation}
and note that for a neutral system the unitary transformed Pseudomomentum (\ref{Kprime}) coincides with CM momentum, $\mathbf {\hat K}^{\prime} = \mathbf {\hat P}$.

\hskip 1cm
Studying the classical neutral system we found unusual special, superintegrable, closed trajectories for a vanishing Pseudomomentum, $\bf K=0$, see \cite{ET:2013}. These trajectories were related with the appearance of the particular integral and were described analytically. Thus, it seems natural in quantum case to consider the zero modes of Pseudomomentum, $\bf K=0$: does exist something unusual?
Since $\mathbf {\hat K}^{\prime}\ =\ \mathbf {\hat P}$, the zero modes of Pseudomomentum correspond to standing composite particle, the system is at rest, there is no center-of-mass dynamics. Hence, we have to look for $R$-independent solutions of (\ref{biespectral})
\begin{equation}
 \Psi^{\prime}_{{}_{0}}(\mathbf R\,, \boldsymbol \rho)\ =\ \psi_{}(\boldsymbol \rho)\ .
\end{equation}
In this case $\psi_{}(\boldsymbol \rho)$ satisfies the equation for the relative motion
\begin{equation}
\bigg[-\frac{\nabla^2_\rho}{2\,m_r}-\frac{1}{2}\om_q\, \hat {l}_z+\frac{m_r\,\Omega^2_q\, {\rho}^2}{2}-\frac{e^2}{\rho}-E\bigg]\psi(\boldsymbol \rho)=0\ ,
\label{Erhoq}
\end{equation}
where
\begin{equation}
 \Om_q  \ = \ \frac{e\,B}{2\,m_r} \,, \quad\om_q=\frac{e \,B\,|\mu_2-\mu_1|}{m_r}\ . \non
\end{equation}
(cf. (\ref{Erhoec})).
It is easy to rewrite the equation (\ref{Erhoq}) in polar coordinates $\boldsymbol \rho=(\rho,\,\varphi)$ and look for the eigenfunction $\psi$ in a factorized form
\begin{equation}
   \psi(\boldsymbol \rho)\ =\ \zeta(\rho)\,\Phi(\varphi)\ .
\label{psiq}
\end{equation}
It can be seen immediately, that
\[
\Phi(\varphi) \ =\  \text{e}^{i\, s\, \varphi}\ ,
\]
where $s=0, \pm 1,\,\pm 2,\,...$,  is the eigenfunction of relative angular momentum,
\[
   \hat {l}_z  =  -i\,\partial_\varphi\ .
\]
For $B \neq 0$ the solution $\zeta(\rho)$ is assumed in the form
\footnote{For $B=0$ the original problem (\ref{Hq}), (\ref{Erhoq}) is reduced to the planar two-body Coulomb problem, see for discussion \cite{PW:2010} and references therein; the limit $B$ tends to zero is singular.}
\begin{equation}
\zeta(\rho) \ = \ \text{e}^{-\frac{m_r\,\Om_q}{2}\,\rho^2}\rho^{| s|}\,p(\rho)\ ,
\label{zetaq}
\end{equation}
the following equation for $p$ occurs
\begin{equation}
\bigg[-\rho\, \pa^2_\rho+(2\,\Om_q\, m_r \rho^2-1-2\,| s| )\,\pa_\rho+(2\,\Om_q\,(1+|s|)
-2  \,E-s\,\om_q)m_r\,\rho \bigg]\,p\ =\ -\ep\,p\ ,
\label{pq}
\end{equation}
being similar to one found in \cite{Turbiner:1994, Taut:1999}. The total energy $E$ takes some special values, see below, where
\[
\ep \equiv 2\,m_r\,e_1\,e_2=-2\,m_r\,e^2\ ,
\]
is introduced. The equation (\ref{pq}) is strikingly similar to (\ref{pec}) with the only difference that the definition of $\ep$ is opposite in sign in front of $e^2$, see (\ref{pec}). Thus, the physical relevant values of $\ep$ should be negative. The equation (\ref{pec}) can be considered as a spectral problem where $\ep$ plays a role of the spectral parameter while the energy $E$ is fixed. The parameter $\ep$ defines a strength of the Coulomb interaction. By changing variable $\rho \rar \sqrt{2\,\Om_q\, m_r}\ \rho$, Eq.({\ref{pq}}) can be reduced to
\begin{equation}
 T_q\ p \equiv  \bigg[-\rho\, \pa^2_\rho+(\rho^2-1-2\,| s| )\,\pa_\rho+(1+|s|
  -\frac{2  \,E+s\,\om_q}{2\,\Om_q})\rho\bigg]\,p=\ -\frac{\ep}{\sqrt{2\Om_q m_r}} p\ .
\label{P-q}
\end{equation}
Eq. (\ref{P-q}) can be written in terms of $sl_2$ generators (\ref{generators}),
\begin{equation}
     \bigg[ -{\hat J}^0_n{\hat J}^- + {\hat J}^+_n -
   (1+2\,|s|+\frac{n}{2}){\hat J}^-  \bigg]p = -\frac{\ep}{\sqrt{2\,\Om_q\, m_r}}\,p\ ,
\label{Tq}
\end{equation}
(cf. (\ref{Tec})), where
$$ n   \equiv \frac{2\,E+\om_q\,s}{2\,\Om_q} -1-|s|\ ,$$
plays a role of the spin of representation. Then the total energy is equal to
\begin{equation}
  E = \Om_q\,(n+1+|s|)-\frac{\om_q\,s}{2}\ ,
\end{equation}
(cf. (\ref{Erho})).

\hskip 1cm
A hidden algebraic structure occurs (the underlying idea behind quasi-exactly solvability) at nonnegative integer $n$: the $sl_2$ algebra appears in finite-dimensional representation and the problem (\ref{Tq}) possesses $(n+1)$ eigenfunctions $p_n$ in the form of a polynomial of the $n$th power. It implies the quantization of $\ep$ as well as the dimensionless parameter
\begin{equation}
 \la \equiv \frac{B_0}{B}\quad ,\quad B_0\ =\ 4\,m_r^2\,e^3        \ ,
\label{lambda}
\end{equation}
which we introduce for convenience, here $B_0$ is a characteristic magnetic field (cf. (\ref{lambdaec})).

\hskip 1cm
It is clear that the cases $e_c=0$ and $q=0$ are described by the same equation (\ref{Tec}) or (\ref{Tq}) with the same discrete values of $\la_n$\ (!). The difference appears on the level of the spectral parameter $\ep \propto \sqrt{\la}$: it should be taken positive values for  the case $e_c=0$ and negative values for the case $q=0$. Corresponding polynomial eigenfunctions of (\ref{Tq}) can be obtained from those of (\ref{Tec}) by replacing $\om_c \rightarrow 2\,\Om_q$ and $\rho\rightarrow-\rho$ , see e.g. (\ref{n=0ec}) - (\ref{n=8ec}). Furthermore, the polynomial eigenfunctions occur for the same discrete values of magnetic field $b_n = \frac{1}{\la_n}$
but different values of the dimensionfull magnetic field $B$.

\hskip 1cm
It is evident that for the operator $T_q(n)$ at integer $n$ (see (\ref{P-q}), (\ref{Tq})),
\[
  [T_q(n)\ ,\ i_n(\rho)]: {\cal P}_{n} \ \mapsto \ 0\ .
\]
Hence, $i_n(\rho)$ is the particular integral with ${\cal P}_{n}$ as the invariant subspace. The eigenfunctions $p_{n, k}\ , \ k=1,\ldots (n+1)$ of (\ref{P-q}), (\ref{Tq})) are the zero modes of $i_n(\rho)$.
Taking the gauge rotated $i_n(\rho)$ with the factor
$\zeta^{(0)} = \text{e}^{-\frac{m_r\,\Om_q \,\rho^2}{2}}\rho^{|s|}\ $ (see (\ref{zetaq})),
\begin{equation}
\label{INT-Hq}
      \zeta^{(0)}\ i_n (\rho)\ (\zeta^{(0)})^{-1}\ =\ \prod_{j=0}^n (\rho {\cal D}_{\rho} + j) \equiv {\cal I}_n\ (\rho)
      \ ,
\end{equation}
where $${\cal D}_{\rho} = \pa _{\rho} + {m_r\,\Om_q\,\rho} - \frac{|s|}{\rho}\ ,$$ is the covariant derivative, we arrive at
\[
  [{\cal {\hat H}}^{\prime}\ ,\
  {\cal I}_n(\rho)]: {\cal V}_{n} \ \mapsto \ 0\ ,
\]
at $b=b_n$, or equivalently at  $\Om_q=({\Om_q})_n$.
Hence, ${\cal I}_n(\rho)$ is the particular integral with ${\cal V}_{n}\ =\ \zeta^{(0)} {\cal P}_n$ as the invariant subspace.

\hskip 1cm
In classical limit the particular integral $i_n(\rho)$ becomes the classical particular integral
$I_n(i\ \rho p_{\rho})$. The latter becomes the constant of motion on the special periodic trajectories. It manifests a connection between quasi-exact-solvability in quantum mechanics and special trajectories in classical mechanics \cite{Turbiner:2013}.

\hskip 1cm
Let us emphasize the existence of what is called {\it particular} integrals (see e.g. \cite{Turbiner:2013}),
\begin{equation}
\label{Ltilda}
    {\tilde{L}}_z = U {\hat {L}}_z U^{-1} = [{\bf R} \times (\mathbf {\hat P} - e_c \mathbf A_{\boldsymbol \rho})]_z \quad , \quad
    {\tilde{\ell}}_z = U {\hat {\ell}}_z U^{-1} = [{\boldsymbol \rho} \times (\mathbf {\hat p} + e_c \mathbf A_{\bf R})]_z\ .
\end{equation}
where ${\hat {L}}_z$ and ${\hat {\ell}}_z$ are CM and relative angular momenta.
In spite of the fact that, in general, the commutators $[{\cal {\hat H}}\,,\, {\tilde {L}}_z\, ]$ and $[{\cal {\hat H}}\,,\, {\tilde {\ell}}_z\, ]$ do not vanish, for all eigenfunctions
\[
   \Psi_{\{0\}} =  e^{ - i\,e_c\,\mathbf A_{\boldsymbol \rho}\cdot \mathbf R}\ \text{e}^{i\, s\, \varphi}\ \text{e}^{-\frac{m_r\,\Om_q}{2}\,\rho^2}\rho^{| s|}\,p_n(\rho)\ ,
\]
(zero modes of $\mathbf {\hat K}$, (\ref{biespectral})), it is satisfied
\begin{equation}
[{\cal {\hat H}}\,,\, {\tilde {L}}_z\, ]\Psi_{\{0\}} = [{\cal {\hat H}}\,,\, {\tilde {\ell}}_z\, ]\Psi_{\{0\}} = 0\ ,
\label{ComK}
\end{equation}
where
\begin{equation}
{\cal {\hat H}}\,\Psi_{\{0\}}\ =\ (\Om_q\,(n+1+|s|)-\frac{\om_q\,s}{2})\,\Psi_{\{0\}}\ ,
\quad{\tilde {L}}_z\,\Psi_{\{0\}}\ =\ 0\,,\quad {\tilde {\ell}}_z\, \Psi_{\{0\}}\ =\
s\,\Psi_{\{0\}}\ .
\label{ComKL}
\end{equation}
It is worth noting the zero modes of the Pseudomomentum, $\Psi_{\{0\}}$, are characterized by zero {\it modified} angular momentum ${\tilde {L}}_z$. It is true for a certain values of a magnetic field $B$.

\hskip 1cm
There exist four global integrals ${\cal {\hat H}},\,\mathbf {\hat K},\,{\hat {L}}^{\rm T}_z$, but the system is not completely-integrable since $[\mathbf {\hat K},\,{\hat {L}}^{\rm T}_z]\neq 0$ (see (\ref{AlgebraInt})). However, over the (sub)space of zero modes of the Pseudomomentum $\Psi_{\{0\}}$ they commute, $[\mathbf {\hat K},\,{\hat {L}}^{\rm T}_z]\Psi_{\{0\}} = 0$. Therefore the system is completely-integrable over the (sub)space of zero modes!
Besides that in the (sub)space of zero modes of the Pseudomomentum, $\Psi_{\{0\}}$, there exist three particular integrals ${\tilde {L}}_z, {\tilde {\ell}}_z, {\cal I}_n(\rho)$ for a certain values of a magnetic field. Hence, the system is maximally super-integrable over the space of zero modes. It is in a complete agreement with the study of the classical case related to special trajectories for which the system possesses a particular, maximal superintegrability.
Perhaps, it has to be emphasized that in the Born-Oppenheimer approximations of the zero order
(say, $m_2 \rar \infty, m_r \rar m_1$) at $B \neq 0$ all conclusions remain valid.

\section*{Conclusions}

A general quantum system of two Coulomb charges on a plane subject to a constant, uniform magnetic field perpendicular to the plane is integrable with Pseudomomentum $K_{x,y}$ and total angular momentum $L_z^T$ as global integrals. No more outstanding properties of a general system we were able to find. It is in contrast to a general classical two Coulomb charge system in a magnetic field which exhibits special superintegrable, periodic trajectories and the existence of the particular integrals which become constant of motion on these trajectories.
However, two particular, physically important quantum systems, $e_c=0$ and $q=0$ at rest (the center-of-mass momentum is zero), reveal a number of outstanding properties. These properties become the most visible when the center-of-mass coordinates are introduced, and parameterized by double polar coordinates in CMS $(R, \phi)$ and relative $(\rho, \varphi)$ coordinate systems, and in addition the Hamiltonian and all integrals are unitary transformed with factor (\ref{U}).

\hskip 1cm
For {\it arbitrary magnetic field}:
(i) eigenfunctions are factorizable (up to the factor (\ref{U})), all factors except the $\rho$-dependable are found explicitly, they have definite relative angular momentum (relative magnetic quantum number),
(ii) dynamics in $\rho$-direction is the {\it same} for both systems, it corresponds to the funnel-type potential (see (\ref{Erhoec}) and (\ref{Erhoq})) and it is characterized by the hidden $sl(2)$ algebra;

while at some {\it discrete values of magnetic fields},

(iii) particular integral(s) occur,
(iv) the hidden $sl(2)$ algebra emerges in finite-dimensional representation, thus, the system becomes quasi-exactly-solvable (of the second type, see \cite{Turbiner:1988}, \cite{Turbiner:1994}) and
(v) a finite number of polynomial eigenfunctions in $\rho$ appear, they marked by extra quantum number(s). Nine families of such analytic eigenfunctions are presented in explicit analytic form.

\hskip 1cm
Quantum system at $e_c=0$ in a magnetic field is completely-integrable (there exist four mutually commuting global integrals) and super-integrable (there exists one extra global integral). At certain discrete values of a magnetic field we are able to find one extra particular integral (\ref{INT}).
Quantum system at $q=0$ in a magnetic field is not completely-integrable (there exist four mutually non-commuting global integrals, in general). However, it becomes completely-integrable over the (sub)space of zero modes of the Pseudomomentum $\Psi_{\{0\}}$ (see (\ref{ComK}), (\ref{ComKL})).
At certain discrete values of a magnetic field we are able to find three extra particular integrals: unitary rotated CM and relative angular momenta ${\hat {L}}_z, {\hat {\ell}}_z$, respectively, and ${\cal I}_n(\rho)$, (\ref{INT}).

\hskip 1cm
It is worth mentioning that a study of the quantum system at $e_c=0$ or at $q=0$ at rest at arbitrary magnetic field is reduced to the study of the dynamics in the relative distance $\rho$ direction. It is a sufficiently simple one-dimensional quantal problem with funnel-type potential. It is explored elsewhere \cite{ET-var}.

\begin{acknowledgments}
  The authors are grateful to J. C. L\'opez Vieyra and, A.V.T. thanks G.~Korchemsky and P.~Winternitz for their interest in the present work and helpful discussions.
  This work was supported in part by the University Program FENOMEC, by the PAPIIT
  grant {\bf IN109512} and CONACyT grant {\bf 166189}~(Mexico).
\end{acknowledgments}



\begin{center}
\textbf{Appendix. Algebra $sl_2$}
\end{center}

The algebra $sl_2$ is realized by the first-order differential operators
\begin{equation}
\begin{aligned}
&     {\hat J}_n^+  =  \rho^2 \pa_\rho-n\,\rho    \,,
\\ &  {\hat J}_n^0  =  \rho \, \pa_\rho-\frac{n}{2}    \,,
\\ &  {\hat J}^-  =  \pa_\rho    \ ,
\label{generators}
\end{aligned}
\end{equation}
where $\pa_{\rho} \equiv \frac{d}{d\rho}$, $n$ is a parameter, see e.g. \cite{Turbiner:1988}. 
These operators are the generating elements of the M\"obius transformation. If $n$ is a nonnegative integer, the algebra $sl_2$ possesses $(n+1)$-dimensional irreducible representation realized in polynomials
of degree not higher than $n$,
\begin{equation}
\label{Pn}
{\cal P}_{n+1}\ =\ \langle  1,\rho,\rho^2,...,\rho^n \rangle   \ .
\end{equation}


\begin{thebibliography}{99}

\bibitem{LL}
        L.D.~Landau and E.M.~Lifshitz,\\
        \textit{ Quantum Mechanics, Non-relativistic Theory {\rm (}Course of Theoretical Physics {\rm vol 3)}},
        {3rd edn (Oxford: Pergamon Press)}, 1977

\bibitem{Turbiner:2013}
         A.V.~Turbiner,\\
         \textit{Particular Integrability and (Quasi)-exact-solvability}\\
         {\it Journal of Physics \bf A45} (2013) 025203 (9pp)\\
         {\tt math-ph arXiv:1206.2907}

\bibitem{ET:2013}
         M.A.~Escobar-Ruiz and A.V.~Turbiner,\\
         \textit{Two charges on a plane in a magnetic field: special trajectories}\\
         {\em Journal of Math Physics \bf 54} (2013) 022901

\bibitem{Kohn:1961}
        W.~Kohn,\\
        \textit{ Cyclotron Resonance and de Haas-van Alphen Oscillations of an Interacting Electron Gas},\\
        {\em Phys. Rev. \bf 123} (1961) 1242–1244

\bibitem{Taut:2000}
         M.~Taut,\\
         \textit{Special analytic solutions of the Schr\"odinger equation for two and three electrons in a magnetic field and ad hoc generalization to $N$ particles},\\
         {\em Journal of Physics: Condens. Matter \bf 12} (2000) 3689 - 3710

\bibitem{Turbiner:1994}
         A.V.~Turbiner,\\
         \textit{Two electrons in an external oscillator potential: hidden algebraic
            structure},\\
         {\it Phys.Rev. \bf A50} (1994) 5335-5337

\bibitem{Turbiner:1988}
          A.V.~Turbiner, \\
          \textit{Quasi-Exactly-Solvable Problems and the $SL(2,R)$ Group}, \\
         {\it Comm.Math.Phys. \bf 118}, 467-474 (1988)

\bibitem{Laughlin:1983}
         R.~Laughlin,\\
         \textit{Quantized motion of three two-dimensional electrons in a strong magnetic field},\\
         {\em Phys. Rev. \bf 27} (1983) 3383 - 3389

\newpage

\bibitem{Taut:1999}
         M.~Taut,\\
         \textit{Two particles with opposite charge in a homogeneous magnetic field: particular analytic solutions of the two-dimensional Schr$\ddot{o}$dinger equation},\\
         {\em Journal of Physics \bf A32} (1999) 5509 - 5515
         
\bibitem{PW:2010} 
         S.~Post, P.~Winternitz,\\
         \textit{An infinite family of superintegrable deformations of
          the Coulomb potential},\\
         {\em Journal of Physics \bf A43} (2010) 222001
         
\bibitem{ET-var} 
         M.A.~Escobar-Ruiz and A.V.~Turbiner,\\
         \textit{Two charges on a plane in a magnetic field: an approximate solution}\\
         (in preparation)
        

\end{thebibliography}
\end{document}